\documentstyle[12pt]{article}

\def\hybrid{\topmargin 0pt      \oddsidemargin 0pt
         \headheight 0pt \headsep 0pt
         \voffset=-0.5cm
         \textwidth 6.25in       
         \textheight 9.5in       
         \marginparwidth 0.0in
         \parskip 5pt plus 1pt   \jot = 1.5ex}
\catcode`\@=11
\def\marginnote#1{}

\newcount\hour
\newcount\minute
\newtoks\amorpm
\hour=\time\divide\hour by60
\minute=\time{\multiply\hour by60 \global\advance\minute by-\hour}
\edef\standardtime{{\ifnum\hour<12 \global\amorpm={am}%
         \else\global\amorpm={pm}\advance\hour by-12 \fi
         \ifnum\hour=0 \hour=12 \fi
         \number\hour:\ifnum\minute<10 0\fi\number\minute\the\amorpm}}
\edef\militarytime{\number\hour:\ifnum\minute<10 0\fi\number\minute}

\def\draftlabel#1{{\@bsphack\if@filesw {\let\thepage\relax
    \xdef\@gtempa{\write\@auxout{\string
       \newlabel{#1}{{\@currentlabel}{\thepage}}}}}\@gtempa
    \if@nobreak \ifvmode\nobreak\fi\fi\fi\@esphack}
         \gdef\@eqnlabel{#1}}
\def\@eqnlabel{}
\def\@vacuum{}
\def\draftmarginnote#1{\marginpar{\raggedright\scriptsize\tt#1}}

\def\draftlabel#1{{\@bsphack\if@filesw {\let\thepage\relax
    \xdef\@gtempa{\write\@auxout{\string
       \newlabel{#1}{{\@currentlabel}{\thepage}}}}}\@gtempa
    \if@nobreak \ifvmode\nobreak\fi\fi\fi\@esphack}
         \gdef\@eqnlabel{#1}}
\def\@eqnlabel{}
\def\@vacuum{}
\def\draftmarginnote#1{\marginpar{\raggedright\scriptsize\tt#1}}

\def\draft{\oddsidemargin -.5truein
         \def\@oddfoot{\sl preliminary draft \hfil
         \rm\thepage\hfil\sl\today\quad\militarytime}
         \let\@evenfoot\@oddfoot \overfullrule 3pt
         \let\label=\draftlabel
         \let\marginnote=\draftmarginnote
    \def\@eqnnum{(\theequation)\rlap{\kern\marginparsep\tt\@eqnlabel}%
\global\let\@eqnlabel\@vacuum}  }

\def\numberbysection{\@addtoreset{equation}{section}
         \def\theequation{\thesection.\arabic{equation}}}

\def\underline#1{\relax\ifmmode\@@underline#1\else
         $\@@underline{\hbox{#1}}$\relax\fi}

\def\titlepage{\@restonecolfalse\if@twocolumn\@restonecoltrue\onecolumn
      \else \newpage \fi \thispagestyle{empty}\c@page\z@
         \def\thefootnote{\fnsymbol{footnote}} }

\def\endtitlepage{\if@restonecol\twocolumn \else  \fi
         \def\thefootnote{\arabic{footnote}}
         \setcounter{footnote}{0}}  
\relax

\numberbysection
\hybrid

\def\beq{\begin{equation}}
\def\eeq{\end{equation}}
\def\p{\partial}

\begin{document}
\begin{titlepage}

\title{ Conformal Maps
 and 
integrable hierarchies
}

\author{P.B.Wiegmann \thanks{James Franck Institute and and Enrico Fermi
Institute
of the University of Chicago, 5640 S.Ellis Avenue, Chicago, IL 60637, USA and
Landau Institute for Theoretical Physics}
  \and A.Zabrodin
\thanks{Joint Institute of Chemical Physics, Kosygina str. 4, 117334,
Moscow, Russia
  and ITEP, 117259, Moscow, Russia}}

\date{September 1999}
\maketitle

\begin{abstract}
We show that conformal maps of simply connected domains  with an analytic
boundary to a unit disk have an intimate relation to the dispersionless 2D
Toda integrable
hierarchy. The maps are determined by a particular solution  to the
hierarchy singled out
by  the conditions  known as
``string equations". The same  hierarchy  locally
solves the  2D inverse potential  problem, i.e.,  reconstruction of the
domain out of a
set of its harmonic moments. This is the same solution which is known to
describe 2D
gravity coupled to $c=1$ matter. We also introduce a concept of the
$\tau$-function for
analytic curves.
\end{abstract}

\vfill

\end{titlepage}
\section{Introduction}
By the  Riemann mapping theorem, any simply connected domain on the complex
plane with a
boundary consisting of at least two points  can be conformally mapped onto
the unit disk.
However, the theorem  does not say how to construct the map. We argue that
the map is explicitly given by a solution of a dispersionless
integrable hierarchy which is a multi-dimensional extension of
the hierarchies of hydrodynamic type \cite{hydro,kr2}.

In this paper, we only consider
simply connected domains bounded by a simple analytic
curve.  We show that in this case conformal maps are given by a particular
solution of the
dispersionless 2D Toda  hierarchy (see \cite{kr2,Ta1} and references therein).
Surprisingly, this solution obeys (and is selected by) the string equation
familiar in
topological gravity and matrix models of 2D gravity
\cite{gravity1,2matrix}.

One may characterize a closed analytic curve by a set of  harmonic moments
of, say the
exterior of the domain surrounded by the curve and its area:
\beq
\label{h1}
C_k=-\displaystyle{\int \!\! \int_{\mbox{{\small exterior}}}}z^{-k}dxdy\,,
\;\;\;\;\;\; k\geq 1 ;\;\;\;\;\;C_{0}=\displaystyle{\int \!\!
\int_{\mbox{{\small
interior}}}}dxdy\eeq
Here $z=x+iy$ and it is implied that the point $z=0$ is inside the domain,
whereas the
point
$z=\infty$ is outside. The integrals at $k=1,2$ are assumed to be properly
regularized. We assume this (infinite) set to be given and  address two
problems:

(1) To find the harmonic moments of the interior
\beq\label{h2}
C_{-k}=\displaystyle{\int \!\! \int_{\mbox{{\small interior}}}}z^{k}dxdy\,,
\;\;\;\;\;\; k\geq 1,
\eeq
and to reconstruct the form of the domain.
This is the 2D inverse potential problem (see e.g. \cite{Brodsky}).
It is known that the solution, if exists,
is unique \cite{convex} if the domain is star-like. It is not
known whether this is true for arbitrary domains with a smooth boundary.
However, if the boundary is smooth, then a small change of moments
uniquely determines a
new domain (also with a smooth boundary).
The integrable hierarchy  gives the solution to this problem through the
$\tau$-function,
thus suggesting a new concept -- the $\tau$-function of the curve.

(2) To construct an invertible conformal map of the
  exterior of the unit circle
onto the exterior of the domain.

Explicitly, the area and the set of moments
of the exterior (and their complex
conjugate) are identified with the times of the
dispersionless 2D Toda hierarchy as follows:
\begin{equation}\label{11}
t\equiv t_{0}=\frac{C_0}{\pi},\;\;\;\;\;
t_k=\frac{C_k}{\pi k},\;\;\;\;\;\bar
t_k=\frac{\bar C_k}{\pi k},\;\;\;\;\; k\geq 1.
\end{equation}
These parameters should be treated as independent
variables.
We prove that the derivatives
of the harmonic moments of the interior $C_{-k}$
with respect to the $t_j$, $\bar t_j$ enjoy the symmetry
  $$\frac{\p C_{-k}}{\p t_j}=\frac{ \p C_{-j}}{\p t_k},\;\;\;\;\;
  \frac{ \p C_{-k}}{\p \bar t_j}=
\frac{\p \bar C_{-j}}{\p t_k}.$$
Therefore, the moments of the interior
can be expressed as derivatives of a single real function which appears to
be the
logarithm of the $\tau$-function
$\tau(t,\{t_k\},\{\bar t_k\})$ of the dispersionless 2D Toda hierarchy:
\begin{equation}\label{12}
v_k \equiv \frac{C_{-k}}{\pi}=
  \frac{\p \,\mbox{log}\,\tau}{\p t_k}\,,
\;\;\;\;\;\;
\bar v_k \equiv \frac{\bar C_{-k}}{\pi}=
  \frac{\p \,\mbox{log}\, \tau}{\p \bar t_k}\,,
\;\;\;\;\;\; k\geq 1
\end{equation}
The equation which determines  the $\tau$-function (the dispersionless
Hirota equation)
and therefore the moments of the interior is given in
Sec.\,5. Supplemented by a proper initial condition provided by the string
equation,
it solves the inverse moments
problem for small deformations of a simply connected domain
with analytic boundary.
The $\tau$-function at $t_k=0,\;k\geq 3$ is given in the Appendix.

In its turn, the conformal map is determined by the dispersionless limit of
the Lax-Sato
equations of the Toda hierarchy. Let $z(w)$
be a univalent function that provides an
invertible conformal map of the exterior of
the unit circle $|w|> 1$ to the exterior of the domain.
Let us represent it by a Laurent series
\beq
\label{z}
z(w)=rw+\sum_{j=0}^{\infty}u_jw^{-j}
\eeq
The conditions that $w=\infty$ is mapped to $z=\infty$ and $r$
is real fix the map, so
the potentials $r,\,u_j$
are uniquely determined by the domain.
The potentials of the conformal map obey an infinite set of
differential equations with respect to the times (harmonic moments).
They  are evolution Lax-Sato equations of the hierarchy.
The series $z(w)$ is identified with
the Lax function.
The Lax-Sato equations are derived in Sec.4.

The coefficients of the conformal map have an equivalent
description in terms of the
$\tau$-function: the inverse map $w(z)$ is explicitly given by the formula
\beq
\label{i}
\mbox{log}\,w=\mbox{log}\,z-\left (
\frac{1}{2}\frac{\p^2}{\p t^2}+\sum_{k\geq 1}
\frac{z^{-k}}{k}\frac{\p^2}{\p t\p t_k}\right )
\,\mbox{log}\,\tau
\eeq

Two comments are in order.
Many equations of Secs.\,3--6 can be found in the
literature on dispersionless
hierarchies \cite{kr2,Ta1,GiKo1,KC},
2D gravity \cite{gravity1,2matrix} and
topological field theories \cite{AoKo,Du2}
without reference to
conformal maps and
inverse potential problem; we adopt them and their proofs for
conformal maps. This work is
the development of our previous joint work \cite{x}
with Mark Mineev-Weinstein, where some of the results  described
below were applied to the
problems of  2D interface dynamics.

\section{Analytic curves  and  the Schwarz function}

  Let  a closed analytic curve be the boundary of a simply connected
domain on  the complex
plane with coordinates $z=x+iy$, $\bar z =x-iy$.
The equation of the curve can be written in the form
\beq
\label{c1}
\bar z=S(z)
\eeq
Thus the curve determines the function $S(z)$ which is defined also outside
the curve.
Moreover it is an analytic function within some domain containing the
curve. We call the
function
$S(z)$ the {\it Schwarz function} of the curve (see e.g. \cite{Davies}). Its
  singularities encode  the information about the curve.

The Schwarz function may not be an arbitrary function of $z$.
It  obeys the unitarity condition: as is seen from (\ref{c1}), its complex
conjugate function\footnote{A comment on the notation:
given a Laurent series
$f(z)=\sum_j f_j z^j$, we set
$\bar f(z)=\sum_j \bar f_j z^j$.
}
is equal to the inverse function, i.e.,
\beq
\label{c2}
\bar S(S(z))=z.
\eeq

Being an analytic function in a strip containing the curve, the Schwarz
function can be
decomposed into a sum of  two functions $S(z)=S^{(+)}(z)+S^{(-)}(z)$: one,
$S^{(+)}$, is
holomorphic in the  interior of the domain, the second, $S^{(-)}$, is
holomorphic in the
exterior. Their expansions around $z=0$ and $z=\infty$ determine the
harmonic moments
  in the exterior (\ref{h1}) and interior  (\ref{h2}) respectively:
\beq
\label{h4}
S^{(+)}(z)=\frac{1}{\pi}\sum_{k=1}^{\infty}
C_{ k}z^{ k-1},\;\;\;\;\;S^{(-)}(z)=\frac{1}{\pi}\sum_{k=0}^{\infty}
C_{ -k}z^{ -k-1}
\eeq
This follows from the contour integral representation of the moments
\beq
\label{h21}
C_k =\frac{1}{2i}\oint_{\mbox{\small curve}}z^{-k} S(z) dz\,,
\;\;\;\;\;\;-\infty <k<\infty
\eeq
obtained from (\ref{h1},\,\ref{h2}) with the help of the Green formula.

The Schwarz function is closely related to conformal maps.
Consider the conformal map (\ref{z}) of the exterior of the unit circle to
the exterior of
the domain. The inverse map
sends a  point $z$ to a point $w(z)$. Let us invert this
point with respect to the circle, $w \to (\bar w)^{-1}$ and map
it back. (It is known that the conformal can be extended into some domain
through the analytic
curve. We assume that the inverted point belongs to it.) This operation
is carried out
by the conjugate Schwarz function
$\bar S(\bar z) =z((\bar w)^{-1})$ or
$S(z)=\bar z(w^{-1})$. Obviously,  if $z$ belongs to the curve,
then $S(z)=\bar z$.
Therefore, we can write $S(z(w))=
\bar z(w^{-1})$.

To summarize, we have two formal
Laurent series  (see (\ref{11},\,\ref{12})
for the notation):
\beq
\label{lg7}
S(z)
=\sum_{k=1}^\infty
kt_{k}z^{k-1}+\frac{t}{z}+\sum_{k=1}^\infty v_kz^{-k-1}
\eeq
\beq
\label{lg17}
\bar S( z)
=\sum_{k=1}^\infty
k\bar t_{k} z^{k-1}+\frac{t}{ z}+\sum_{k=1}^\infty \bar v_k z^{-k-1}
\eeq
They are connected by the
unitarity condition (\ref{c2}). The latter is resolved by the conformal maps
\beq
\label{h10}
z(w)=\bar S(\bar z(w^{-1}))
\eeq
\beq\label{h101}
\bar z (w^{-1})=S(z(w))
\eeq
In their turn, the conformal maps are given by the
series\footnote{To avoid confusion, let us stress that the functions $z(w)$
and $\bar
z(w^{-1})$ are complex conjugate only on the curve, i.e., at $|w|=1$.}
\beq
\label{c41}
  z(w)=
rw+\sum_{j=0}^{\infty}u_jw^{-j}
\eeq
\beq
\label{c4}
\bar z(w^{-1})=
rw^{-1}+\sum_{j=0}^{\infty}\bar u_jw^{j}
\eeq

They and the unitarity condition (\ref{h10})  establish relations between
harmonic
moments of the exterior $t_k$,
harmonic moments of the interior $v_k$ and the
coefficients  of the conformal map $u_j$.
Below we find an infinite set of
differential equations, which determine
the evolution of the potentials and
the moments of
the interior  in times -- moments of the interior.

\section{Symplectic structure of conformal maps and generating function}

Deformations of the domain and, therefore, of the conformal map reveal the
symplectic structure. In this section we show that the pairs $\mbox{log}
w,\,t$ and
$z(w,t),\,\bar z(w^{-1},t)$ are canonical (see also \cite{x}):
\begin{equation}\label{pb}
\{z(w,t),\,\bar z(w^{-1},t)\}=1
\end{equation}
where $\{,\} $ is the Poisson bracket with respect to $w$ and the area $t$
(all
moments of the exterior $t_k,\, \bar t_k$ are kept fixed).  For any two
functions
$f(w,t)$,
$g(w,t)$ the Poisson bracket is defined by
\beq
\label{PB}
\{f,\,g\}=w\frac{\p f}{\p w}\frac{\p g}{\p t}-w\frac{\p g}{\p w}\frac{\p
f}{\p t}
\eeq

We refer to (\ref{pb}) as the string equation\footnote{See Sec.\,7
for the history of this equation and references.}.
This  equation suggests that
deformations of conformal maps form a group with
respect to the composition.
To prove it, let us rewrite the l.h.s.
in two equivalent forms with the help of
(\ref{h10}), (\ref{h101}). First, let us compute the derivatives  of $\bar
z(w^{-1},t)$
in (\ref{pb})  treating it as the composition $S(z(w,t),t)$ as is in
(\ref{h101}):
\begin{equation}
\label{c5}
\frac{\partial \bar z(w^{-1},t)}{\partial t}=
  \frac{\partial S(z,t)}{\partial t}+
  \frac{\partial S(z,t)}{\partial z}
\frac{\partial z(w,t)}{\partial t},\;\;\;\;\;\;
\frac{\partial  \bar z(w^{-1},t)}{\partial w}=
   \frac{\partial S(z,t)}{\partial z}
\frac{\partial z(w,t)}{\partial w}
\end{equation}
As a result, we obtain:
\begin{equation}\label{pb1}
\{z(w,t),\,\bar z(w^{-1},t)\}=w\, \frac{\partial z(w,t)}{\partial
w}\frac{\partial
S(z,t)}{\partial t}
\end{equation}
Similarly, treating $z(w,t)$ in (\ref{pb}) as the composition
$\bar S(\bar z(w^{-1},t),t)$ we get
\begin{equation}\label{pb2}
\{z(w,t),\,\bar z(w^{-1},t)\}=-w\,\frac{\partial\bar z(w^{-1},t)}{\partial
w}\frac{\partial
\bar S(\bar z,t)}{\partial t} \,.
\end{equation}
In the r.h.s. of these equations the derivatives in  $t$ is taken at fixed
$z$ or $\bar z$
and then understood  as  functions of $w$.

Now, using the series (\ref{lg7}) and (\ref{c41}), we conclude that the
r.h.s. of
(\ref{pb1}) is 1 plus positive powers in $w$. However,
the series (\ref{lg17}) and (\ref{c4}) tell  that the r.h.s. of
(\ref{pb2}) is 1 plus negative powers in $w$.
This prompts us to (\ref{pb}).

The rest follows from the symplectic structure by
treating deformations of the conformal map
(\ref{c41}) along the lines of the multi-time Hamilton-Jacobi formalism
\cite{kr2,Du1}.

  Let us introduce the generating function
$\Omega (z,t)$ of the canonical transformation
$\,(\mbox{log}w,\,t)\to (z,\, S)$. Its
differential, $d\Omega=Sdz+\mbox{log}\, w dt$,
implies that
\begin{equation}
\label{HJ2}
S=
\frac{\partial \Omega (z,t)}{\partial z}\,,
\;\;\;\;\;\;
\mbox{log}\,w=
\frac{\partial \Omega(z,t)}{\partial t}
\end{equation}
Using the Laurent series for the Schwarz function (\ref{lg7}), we  get
\beq
\label{HJ4}
\Omega (z,t)=\sum_{k=1}^{\infty} t_kz^k
+t\,\mbox{log}\,z -\frac{1}{2}v_0(t)-
\sum_{k=1}^{\infty}\frac{v_k (t)}{k}z^{-k},
\eeq
where $v_0$ obeys
\beq
\label{v}
\p_t v_0=2\,\mbox{log} \,r
\eeq
The integration constant $v_0$ can be interpreted as  a logarithmic
moment (see (\ref{v0}) below).

As the Schwarz function, the generating function
can be represented as the sum
of functions, whose derivatives,
$S^{(\pm)}(z)=\p_z\Omega^{(\pm)}(z)$,
are analytic in the exterior and the interior of the domain respectively:
$\Omega (z)=\Omega^{(+)}(z)+
\Omega^{(-)}(z)-v_0/2$.
They have a simple electrostatic interpretation. Say,
$\Omega^{(+)}(z)$ is the (complex) 2D Coulomb
potential in the interior of the domain created by a homogeneously
distributed charge in
the exterior. In its turn, the $\Omega^{(-)}(z)$ is the complex Coulomb
potential
in the exterior created by a homogeneous charge in the interior:
\beq
\label{lg8}
\Omega^{(+)}(z)=\frac{1}{\pi}
\int \!\!\!\int_{\mbox{{\small exterior} } }
\mbox{log}\Bigl ( 1-\frac{z}{z'} \Bigr ) dx' dy'
= \frac{1}{\pi}\sum_{k=1}^{\infty} \frac{C_k}{k}z^k
\eeq
\beq
\label{lg81}
\Omega^{(-)}(z)=\! \frac{1}{\pi}\!
\int \!\!\!\int_{\mbox{{\small interior} } }
\!\!\! \mbox{log}\Bigl ( z\!-\!{z'} \Bigr ) dx' dy' \!
=\!\frac{C_0}{\pi}\,\mbox{log}\,z \! -\!\frac{1}{\pi}\!\sum_{k=1}^{\infty}
\frac{C_{-k}}{k}z^{-k}
\eeq

The real and imaginary
parts of the generating function treated as a function
of a point $z$ on a curve have a clear geometric interpretation.
The real part is half of the squared distance from the
origin to the point $z$ on a curve while
$\frac{1}{2}\,\Im m\,(\Omega(z,t)-\Omega(z_1,t))$
is the area $A(z)$ (counted modulo $\pi t$) of the
interior domain bounded by the ray
$\varphi =\mbox{arg}\,z$ and
a reference ray $\varphi =\mbox{arg}\,z_1$. So,
if $z$ is on a curve, we can write
\beq
\label{reim}
\Omega (z) =\frac{1}{2}|z|^2+2iA(z)\,,
\eeq
where the right hand
side is defined up to a purely imaginary constant.
Indeed, integrating by parts $\Omega (z)=\int^z S(\zeta )d\zeta =
\frac{1}{2}zS(z)+\frac{1}{2}\int^z (S(\zeta )d\zeta -\zeta dS(\zeta ))$
and taking into account that
$dA(z)=\frac{1}{4i}(\bar z dz -zd\bar z)$ and
$S(z)=\bar z$ on a curve,
we obtain (\ref{reim}).

It follows from (\ref{reim}) that if $z$ belongs
to a curve, then the variation of the
$\Omega (z)$ with respect to the {\it real} parameters,
\beq
\label{Om1}
\delta \Omega =\p_t \Omega \delta t+
\sum _{k=1}^{\infty}\left [
(\p_{t_k}\Omega +\p_{\bar t_k}\Omega )\,{\cal R}e\,\delta t_k +
i
(\p_{t_k}\Omega -\p_{\bar t_k}\Omega )\,{\cal I}m\,\delta t_k
\right ]
\eeq
is purely imaginary.
The proof is simple.
Let us analytically continue the
equality $\Omega (z)+\bar \Omega (\bar z)=|z|^2$
(the real part of (\ref{reim})) away from the curve,
\beq
\label{a5}
\Omega (z)+\bar \Omega (S(z))=zS(z)
\eeq
and take the partial derivative with respect to $t_j$
at a fixed $z$.
Note that the $t_j$-dependence
of the second term comes from both $\bar \Omega$ and $S$.
We get:
$$
\left.
\frac{\p \Omega (z)}{\p t_j}+
\bar S(S(z))
\frac{\p S(z)}{\p t_j}+
\frac{\p \bar \Omega (\zeta )}{\p t_j}\right |_{\zeta =S(z)}
=z\, \frac{\p S(z)}{\p t_j}
$$
Applying the unitarity condition (\ref{c2}),
we observe that the second term in the left hand side
is equal to the term in the right hand side, so they cancel.
Restricting the equality to the curve again,
we conclude that
\beq
\label{a6}
\frac{\p \Omega (z)}{\p t_j}+
\frac{\p \bar \Omega (\bar z)}{\p t_j} =0
\eeq
whence
all terms of the infinite sum in
the right hand side of (\ref{Om1}) are
imaginary. Similarly, for the derivative
with respect to $t$ (\ref{a6}) reads
$\p_t \Omega +\overline{\p_t \Omega}=0$
(since $t$ is real), so the first term in (\ref{Om1})
is also imaginary.

As is mentioned above, $\Omega$ is defined up
to a purely imaginary $z$-independent term (which may depend
on the $t_j$).
However, its variation with respect to
{\it real} parameters, as in (\ref{Om1}), is also
purely imaginary.
Taking into account (\ref{v}), we fix it by the
condition that $v_0$ is real.
From (\ref{lg8}), (\ref{lg81}) and (\ref{reim}) it easily
follows that
\beq
\label{v0}
v_0=\frac{1}{\pi}\int_{\mbox{{\small interior} } }
\mbox{log}\, |z|^2 dx dy
\eeq
(the logarithmic moment). To prove this formula, one should
continue the equality $v_0 -2{\cal R}e\, \Omega^{(+)}(z)=
2{\cal R}e\, \Omega^{(-)}(z)-|z|^2$ (which is just the real part
of (\ref{reim})) {\it harmonically} away from the boundary
into the interior and evaluate both sides at $z=0$.
\section{Lax-Sato equations for conformal maps}

Let us now vary higher  harmonic moments  $t_k$.  The  differential of the
generating
function changes accordingly \cite{x}:
\begin{equation}
\label{5}
d\Omega =Sdz+\,\mbox{log}\,
  w dt+\sum_{k=1}^\infty(H_k dt_k-\bar H_k d\bar t_k)
\end{equation}
where
\begin{equation}
\label{HJ5}
H_j=\left
(\frac{\partial \Omega}{\partial t_j}\right )_{z},\;\;\;\;\bar H_j=-\left
(\frac{\partial \Omega}{\partial \bar t_j}\right )_{z}
\end{equation}
are Hamiltonians generating  the
higher flows in $t_k, \bar t_k$.
Here the derivatives in $t_k$ are taken at a fixed $z$
that belongs to the curve.
The second equality follows from (\ref{a6}).
The Hamiltonians, being treated as
functions of $z$ and $ t$ obey the integrability condition \cite{kr2}
\begin{equation}
\label{HJ6}
\left ( \frac{\p H_i}{\p t_j}\right )_z
=\left ( \frac{\p H_j}{\p t_i}\right )_z
\end{equation}
Computing the derivative in (\ref{HJ5}), we obtain:
\beq
\label{HJ7}
H_j=z^j -\frac{1}{2}\frac{\p v_0}{\p t_j}
-\sum_{k=1}^{\infty} \frac{\p v_k}{\p t_j}\frac{z^{-k}}{k}
\eeq
and similarly for $\bar H_k$.

The Hamiltonians $H_j$ written in terms of
the canonical variables $w,\,t$
determine the
evolution of the conformal map $z(w)$ and $\bar z(w^{-1})$
with respect to the hierarchical times $t_k$ (harmonic moments).
These are Lax-Sato
equations:
\beq
\label{HJ9a}
\frac{\p z(w)}{\p t_j} =\{H_j, z(w)\}
\eeq
\beq
\label{HJ9b}
\frac{\p \bar z(w^{-1})}{\p t_j} =\{H_j, \bar z (w^{-1})\}
\eeq
Note that these formulas can be
extended to $j=0$ where $H_0 =\,\mbox{log}\, w$
generates the flow $t_0 =t$.
Consistency of these equations
acquires the form of the "zero curvature" condition
$\p _{t_j} H_i
-\p _{t_i}H_j +\{H_i,\,H_j\}=0$.
It is equivalent to  (\ref{HJ6}) rewritten in the variable $w$.

Now we are ready to prove that the Hamiltonians $H_i$ and $\bar H_i$
have the  form
\beq
\label{HJ12}
H_j(w)=\Bigl (z^j(w)\Bigr )_{+} +\frac{1}{2}
\Bigl (z^j(w) \Bigr )_{0}
\eeq
\beq
\label{J4}
\bar H_j(w)=\Bigl (\bar z^j(w^{-1})\Bigr )_{-} +\frac{1}{2}
\Bigl (\bar z^j(w^{-1}) \Bigr )_{0}
\eeq
The symbols $(f(w))_{\pm}$  mean a truncated Laurent series, where
only  terms with positive (negative)
powers of $w$ are kept, $(f(w))_{0}$ is a
constant part ($w^0$) of the series.

To this end, let us differentiate $H_j$ by $w$ and $t$, and express the
result in
terms of derivatives of  $\bar z$ and $\bar S$. We begin with the formula
$$
\frac{\p H_j}{\p w}=\frac{\p z(w)}{\p w}
\frac{\bar z (w^{-1})}{\p t_j}-
\frac{\p z(w)}{\p t_j}\frac{\bar z(w^{-1})}{\p w}
$$
which is a simple consequence of the definition (\ref{HJ5}).
Replacing here $z(w)$ by $\bar S(\bar z)$ as before, we get:
\beq\label{x}\frac{\p H_j}{\p w}=\frac{\p \bar z}{\p
w}\frac{\p \bar S(\bar z)}{\p t_j}
\eeq
Using (\ref{lg17}), we find that $\p_{t_j}\bar S (\bar z)=\sum_{k=1}^\infty
\p_{t_j}\bar v_k \bar z^{-k-1}$ is a regular function of $w$ at $w=0$ and
its Taylor
expansion starts from $w^{2}$. Thus $H_j$ is also a regular function in
$w$. Moreover,
from the (\ref{HJ7}) we find that $H_j$ is a polynomial in $w$ of the
degree $j$. Thus,
being so, we have
$H_j=(z^j)_++(z^j)_0-\frac{1}{2}\p_{t_j}v_0$. To complete the proof, let us
find the
$w^0$-term in the Laurent series
\beq
\label{HJ10}
\frac{\p H_j}{\p t}=
\frac{\p \bar S(z)}{\p t}\frac{\p \bar z(w)}{\p t_j}-
\frac{\p \bar z(w)}{\p t}\frac{\p \bar S(z)}{\p t_j}
\eeq
  It comes from the first term of the r.h.s. of this
expression and, together with (\ref{HJ4}), gives the desired result:
$2(\p_tH_j)_0=2\p_{t_j}\mbox{log} r=\p_t\p_{t_j} v_0=\p_t(z^j)_0$.
The dynamics of the conformal map with respect to $\bar t_k$ can be obtained
from (\ref{HJ9a},\,\ref{HJ9b}) by the comlex conjugatiion.
Note that the Poisson bracket
changes the sign as $w\to \bar w=w^{-1}$ that is just consistent with the
minus sign in
(\ref{HJ5}). Hence $\bar H_j =\bar H_j(w^{-1})$ as is in (\ref{J4}).

The Lax-Sato equations (\ref{HJ9a},\,\ref{HJ9b}) with the
Hamiltonians (\ref{HJ12},\,\ref{J4})
imply that the coefficients of the conformal map obey an infinite
set of non-linear differential equations known as the dispersionless
2D Toda hierarchy.
  The first and the most
familiar
equation of the hierarchy is the long wave limit of the Toda lattice
equation:
  $\p ^2 (r^2)/\p t^2 =
\p ^2 \mbox{log}\, (r^2)/\p t_1 \p \bar t_1$.

We also mention other relations between the conformal map and the
Hamiltonians:
\beq
\label{S4a}
z(w)=H_1 +
\frac{1}{2}\bar t_1+\sum_{k=2}^{\infty}k\bar t_k \bar H_{k-1}
\eeq
\beq
\label{ld1H}
z(w)\bar z(w^{-1}) =t+
2\,{\cal R}e \sum_{k=1}^{\infty} kt_k H_k
\eeq
They can be immediately obtained
from the unitarity condition ({\ref{h10},\,
\ref{h101}) by
comparing the positive and constant parts of the Laurent series in $w$ and
using (\ref{HJ12},\,\ref{J4}). These formulas illustrate an important
relation between
harmonic moments and conformal maps. If all moments $t_k=0$ for $k> N$,
then the
Laurent series  of the conformal map (\ref{z}) is truncated: $u_j=0$ for
$j\geq N$. This,
in particular, gives  another proof of  Sakai's theorem \cite{Sakurai}: if
all but the
first three moments $t,\,t_1,\,t_2$ of the complement of a simply connected
domain on the
complex plane are zero, then the domain is an ellipse.

The Lax-Sato equations (\ref{HJ9a},\,\ref{HJ9b}) supplemented
by the string equation (\ref{pb}) give the complete set of differential
equations for the
potentials $u_j$ of the conformal map as functions of moments $t, t_k, \bar
t_k$. They
uniquely determine small deformations of the map and eventually the curve.

\section{Symmetry of moments and the $\tau$-function of conformal maps and
curves}

The integrable hierarchy suggests a concept of $\tau$-function for  curves
and conformal
maps. The $\tau$-function solves the problem of moments, i.e., restoration
of the moments
of the interior
$v_k$ out of the moments of the exterior $\{t_k\}$ and the area $t$. We
define the
$\tau$-function $\tau(t;\{t_i\},\{\bar t_i\})$ as a real function, which
determines the
moments of the interior by the formulas
\begin{equation}
\label{tau1}
v_k=\frac{\p \, \mbox{log}\,\tau}{\p t_k}\,,\;\;\;\;\;
\bar v_k=\frac{\p \,\mbox{log}\, \tau}{\p \bar t_k}\,,
\;\;\;\;\;
v_0=\frac{\p \,\mbox{log}\, \tau}{\p t}
\end{equation}

The very existence of the $\tau$-function is
  due to the fundamental symmetry of
harmonic moments:
\begin{equation}\label{symmetry}
\frac{\p v_j}{\p t_k}=\frac{\p v_k}{\p t_j},\;\;\;
\;\;\frac{\p v_j}{\p \bar t_k}=\frac{\p\bar
v_k}{\p t_j},\;\;\;\;\;\frac{\p v_0}{\p t_k}=\frac{\p v_k}{\p t}
\end{equation}
which follows \cite{kr2} from the Lax-Sato equations.
We prove this below
following the lines of Ref.\,\cite{Ta1}. To prove the
first symmetry relation, we notice from (\ref{HJ7}) that
$\p_{t_j}{ v_k}$ is the constant ($z^0$) part of the Laurent series
$z^{k+1}\p_z H_j$ in
$z$, i.e., the residue
$\mbox{res}(z^{k}dH_j)$. Then, using the well known property of residues,
we find
that this is equal to the constant part ($w^0$) of the Laurent series
$z^kw\p_w H_j$ in
$w$. Then, using the equation (\ref{HJ12})
we get:
$$\frac{\p v_j}{\p t_k}=(z^kw\frac{\p} {\p w}H_j)_0=(z^kw\frac{\p} {\p
w}(z^j)_+)_0=$$
$$((z^k)_-w\frac{\p} {\p w}(z^j)_+)_0=
-((z^j)_+w\frac{\p} {\p w}(z^k)_-)_0=((z^j)_-w\frac{\p} {\p
w}(z^k)_+)_0=\frac{\p v_k}{\p
t_j}$$
This chain of equalities is due to the  identity $(f_-w\p_w g_+)_0=
  -(g_+w\p_w f_-)_0=
(g_-w\p_w f_+)_0$ for residues of Laurent series. The proof of the other two
symmetry relations is similar.

In fact, the symmetry relations follow from the unitarity
condition (\ref{c2}) for the Schwarz function. To show this,
let us outline another proof which does not rely on existence
of the conformal map. The proof uses only the fact that
the variation of the $\Omega (z)$ with respect to the harmonic
moments $t_j$, if $z$ belongs to the curve, is purely
imaginary (see the end of Sec.\,3).
Using the expansion of $\Omega$ in the series (\ref{HJ4}) and
the second equality in (\ref{HJ5}), one concludes that
\beq
\label{H3}
\oint_{\mbox{{\small curve}}}H_j(z)dH_k(z)=0
\eeq
for all $j,k >0$.
Therefore, we have:
$$
\begin{array}{lll}
&&
\p_{t_j} v_k \,=\,
\displaystyle{\frac{1}{2\pi i}}
\displaystyle{\oint_{\mbox{ {\small curve} } } } z^kdH_j(z) \\&&\\
&=&
\displaystyle{\frac{1}{2\pi i}}
\displaystyle{\oint_{\mbox{{\small curve}}}} H_k(z)dH_j(z) +
\displaystyle{\frac{1}{2\pi i}}
\displaystyle{\oint_{\mbox{{\small curve}}}}
\left (
\displaystyle{\frac{1}{2}}\p _{t_k}v_0 +
\displaystyle{\sum_{l\geq 1}}\p_{t_k} v_l
\, \frac{z^{-l}}{l}\right )dH_j(z) \\&&\\
&=&
\displaystyle{\frac{1}{2\pi i}}
\displaystyle{\oint_{\mbox{{\small curve}}}}
\left (
\displaystyle{\frac{1}{2}}\p _{t_k}v_0 +
\displaystyle{\sum_{l\geq 1}}\p_{t_k} v_l
\, \frac{z^{-l}}{l}\right )
d z^j = \p_{t_k} v_j
\end{array}
$$
and similarly for the other derivatives.

The $\tau$-function determines the Hamiltonians as
functions of $z$. Eq. (\ref{HJ7}) reads
$$H_j=z^j-\left (\frac{1}{2}\frac{\p^2}{\p t\p t_j}+
\sum_{k\geq  1}\frac{z^{-k}}{k}\,\frac{\p^2}{\p
t_j\p t_k}\right )\,\mbox{log}\,\tau $$
The second equation in (\ref{HJ2}) gives the formula
for the inverse conformal map $w(z)$
through the $\tau$-function (see (\ref{i}) in the Introduction).
Formulas for the coefficients $u_j$ of the direct map $z(w)$
through $\mbox{log}\, \tau$ are also available but they have
more complicated structure. The first two are simple:
$r=\frac{1}{2} \exp (\p_{t}^{2}\mbox{log}\,\tau )$,
$u_0=\p^2 \mbox{log}\,\tau / \p t \p t_1$.

The $\tau$-function itself obeys the dispersionless limit of the Hirota
equation (a leading term of the differential Fay identity
\cite{GiKo1},\,\cite{Ta1},\,\cite{KC}):
\begin{equation}
\label{Hirota}
(z\! -\! \zeta)\,\exp \! \left(\sum_{n,m\geq
1}\frac{v_{nm}}{nm}z^{-n}\zeta^{-m}\right)\! =
\! z\exp \! \left(-\sum_{k\geq
1}\frac{v_{0k}}{k}z^{-k}\right)\! -\!\zeta\exp \! \left(-\sum_{k\geq
1}\frac{v_{0k}}{k}\zeta^{-k}\right)
\end{equation}
This is an infinite set  of relations between the second derivatives
$v_{nm}=
\p_{t_n}\p_{t_m}\mbox{log}\tau,\;\;v_{0m}=
\p_{t}\p_{t_m}\mbox{log}\tau$
of the $\tau$-function.
The equations   appear  while expanding the
both sides  of (\ref{Hirota}) in
powers of $z$ and $\zeta$.
In particular, the leading terms as $\zeta \to \infty$ yield
\beq
\label{HH}
z-v_{01}-\sum_{k\geq 1}\frac{v_{1k}}{k}z^{-k}=
 z\exp \! \left(-\sum_{k\geq 1}
\frac{v_{0k}}{k}z^{-k}\right)
\eeq
This is in fact the relation $H_1=re^{H_0} +u_0/2$ between the
Hamiltonians $H_1$ and $H_0 =\mbox{log}\, w$ which follows from
the definition (\ref{HJ12}). Substituting (\ref{HH}) back to
(\ref{Hirota}), one can eliminate $v_{0k}$. The resulting
relations between $v_{mn}$ with $m,n \geq 1$ is exactly the
dispersionless limit of the Hirota equations for the KP
hierarchy from \cite{GiKo1},\,\cite{Ta1},\,\cite{KC}).
Supplemented
by proper initial data, satisfying
the unitarity
condition, equations (\ref{Hirota}) formally solve the local problem of
moments
(or the 2D potential problem),
i.e., reconstruction
of the Laurent expansion of the $S^{(-)}$.

The proof of (\ref{Hirota}) is similar to the
one known in the context of the KP
hierarchy \cite{GiKo1,KC}.
Let
\beq
\label{rwz}
rw=z+\sum_{j=0}^{\infty}p_j z^{-j}
\eeq
be the Laurent expansion of the inverse conformal map
$w(z)$, where $p_j$ are some coefficients.
Multiply both sides by $\zeta ^{-k}z^{k-1}(w)$,
extract the polynomial part in
$w$ and sum over $k\geq 1$. After some rearrangement and using
(\ref{rwz}) for $w(\zeta )$ one gets the relation
$$
r(w(\zeta ) -w(z)) \sum_{k=1}^{\infty} \zeta^{-k-1} (z^k(w))_{+}=
rw(z) \sum_{k=0}^{\infty} \zeta^{-k-1} (z^k(w))_{0}
$$
Using (\ref{HJ12}) and the
identity $\p \mbox{log}\, w(z)/\p z=\sum_{k=0}^{\infty}
z^{-k-1}(z^k(w))_{0}$, we rewrite it in the form
$$
1+2\sum_{n=1}^{\infty}\zeta^{-n}H_n(w(z)) =
\zeta \frac{\p \mbox{log}\, w(\zeta )}{\p \zeta }
\,\frac{ w(\zeta )+w(z)}{ w(\zeta )-w(z)}
$$
Now, plugging the equivalent definition (\ref{HJ7}) of
$H_n$ and integrating this equality with respect to $\zeta$,
one finally arrives at the following important relation:
\beq
\label{prehir}
\sum_{k,n\geq 1}\frac{z^{-k}\zeta^{-n}}{kn}v_{kn}=
\mbox{log}\, \frac{w(z)-w(\zeta )}{z-\zeta }
+\mbox{log}\,r
\eeq
Because of (\ref{i}) this is the same as (\ref{Hirota}).

Let us note, in passing, that the Schwarzian derivative
of the inverse map $w(z)$,
\beq
\label{sd}
T(z)=\frac{w'''(z)}{w'(z)}-\frac{3}{2}\left (
\frac{w''(z)}{w'(z)}\right )^2
\eeq
(the classical stress-energy tensor) admits
a remarkably simple
representation {\it linear} in $\mbox{log}\,\tau$.
Indeed, differentiating both sides
of (\ref{prehir}) with respect to $z$ and $\zeta$,
we get
$$
\sum_{k,n\geq 1}z^{-k-1}\zeta^{-n-1}v_{kn}=
\frac{w'(z)w'(\zeta )}{(w(z)-w(\zeta ))^2}
-\frac{1}{(z-\zeta )^2}
$$
A simple calculation shows that
the limit of the right hand side as $\zeta \to z$ is equal to
$\frac{1}{6}$ of
the Schwarzian derivative (\ref{sd}). Therefore, we obtain:
\beq
\label{Schwder}
T(z)=6\,z^{-2}\sum_{k,n\geq 1}z^{-k-n}
\frac{\p^2 \mbox{log}\,\tau}{\p t_k \p t_n}
\eeq
The Hirota equation itself is thus
a ``splitted" ($\zeta \neq z$) analogue of (\ref{Schwder})

\section{Conformal maps as a reduction of the dispersionless Toda lattice
hierarchy}

The reader familiar with integrable hierarchies of non-linear differential
equations is
able to identify the dynamical system  for conformal maps
(\ref{HJ9a}), (\ref{HJ9b}) with the
dispersionless limit of the Toda lattice hierarchy \cite{kr2,Ta1}. The
latter is related
with the Whitham hierarchy -
the theory of solitons with distinct fast and slow variables. The Whitham
hierarchy
appears after averaging over fast variables (see
\cite{kr2,Du2} and references therein).
  The dispersionless limit emerges as the
genus-zero Whitham hierarchy. Formally, it is a semiclassical limit
$\hbar\to 0$ of  pseudo-differential (or difference) operators. For  the Lax
operator of the 2D Toda lattice,
\beq
\label{L}
L=r(t)e^{\hbar\frac{\p}{\p t}}+
\sum_{j=0}^{\infty}u_j(t)e^{-j\hbar\frac{\p}{\p t }}
\eeq
one should replace the difference operator
$e^{\hbar\frac{\p}{\p t}}$  by
the canonical variable $w$
with the Poisson bracket
$\{\mbox{log}w,t\}=1$. The Lax operator then becomes the Lax function
$L(w)$ given by a
formal  Laurent series in
$w$. The Lax function is identified with the conformal map $z(w)$ (\ref{z}).
The derivatives of the $z(w)$ with respect to the times $t_k$ are given by
(\ref{HJ9a})
  which is nothing else than the dispersionless limit of the
Lax-Sato equation $\hbar \p /\p t_j L =[H_j ,\, L]$, where
$H_j =\bigl (L^j \bigr )_{+}+\frac{1}{2}\bigl (L^j \bigr )_{0}$. (The
coefficient
$1/2$ is due to a particular choice of gauge in (\ref{L}), where the
coefficient
in front of the first term is not fixed to be 1.)
The mathematical theory of the dispersionless hierarchies constrained by
the string
equations has been developed in Refs.\,\cite{kr2,Du1}
  and extended to the Toda hierarchy in Ref.\,\cite{Ta1}. For a
comprehensive review see,
e.g.,\,\cite{Du2}.

In the Toda theory,
there are two sets of independent times, $t_i$ and $\tilde t_i$, and two
sets of
potentials, $u_j$ and $\tilde
u_j $. The identification with
conformal maps requires them to be complex conjugate: $\tilde t_i=\bar
t_i,\;\;\tilde
u_j =\bar u_j$ and the function $r$ to be real.
Under the reality conditions the semiclassical limit of the second Lax
operator,
\beq
\label{L1}
\tilde L=r(t\! -\! \hbar )e^{-\hbar\frac{\p}{\p t}}+
\sum_{j=0}^{\infty}\tilde
u_j(t)e^{j\hbar\frac{\p}{\p t}}
\eeq
is identified with $\bar z(w^{-1})$. The reality condition is consistent
with the
2D Toda hierarchy and selects a class of solutions relevant to
conformal maps.

This class is reduced to a unique solution by imposing the string equation
$\{z,\,\bar
z\}=1$, which in the dispersionful case would be
$[L,\,\bar L]=\hbar$. To clarify the origin of the string equation, one
needs two
more operators. These are the
Orlov-Shulman operators \cite{Or}
\beq
\label{M}
M=\sum_{k=1}^{\infty}kt_kL^k +t +
\sum_{k=1}^{\infty} v_k L^{-k}
\eeq
\beq
\label{barM}
\tilde M=\sum_{k=1}^{\infty}k\tilde t_k\tilde L^k +t +
\sum_{k=1}^{\infty} \tilde v_k \tilde L^{-k}
\eeq
obeying  the conditions $[L,M]=\hbar L\,;\;\;[\tilde L,\tilde
M]=-\hbar\tilde L$. Then the
string equation follows from the relations
\cite{gravity1} $\tilde L=L^{-1}M,\;\;\; \tilde
M=M$.
  From
Secs.\,2,3 it follows that the
dispersionless limit of the operator $L^{-1}M$ enjoys a simple geometric
interpretation:
it is the Schwarz function
$S(z)$ of an analytic curve.

\section{More connections and  equivalences}

{\it Analytic curves and dispersionless hierarchies}. Thus one particular
solution of
the dispersionless 2D Toda hierarchy describes  evolution of the univalent
conformal map
of a domain bounded by a simple
analytic curve to the exterior of the unit
circle. The set of
times of the hierarchy appears to be equivalent to the set of harmonic
moments of the
domain, whereas the conformal map itself is the dispersionless limit of the
Lax operator.

This proposition can be reversed: univalent conformal maps of the
exterior of the unit circle generate a solution
of the dispersionless Toda hierarchy.
The solution is selected by the string equation (\ref{pb}).

Other
solutions of the dispersionless Toda hierarchy are selected by more general
string
equations. The latter are characterized by any two functions $f$ and $g$
forming a
canonical pair: $\{f,g\}=f$. Then the general string equations consistent
with the hierarchy
are $\bar L = f^{-1}(L,M)$, $\bar M = g(L,M)$ (see \cite{Ta1}).
We expect that some of them are also relevant to the conformal maps of
simply connected
domains. It is likely that other types of string equations describe
mappings to or from
domains other then the exterior of the unit circle and also nonunivalent maps.

Moreover, we expect that other integrable hierarchies of the hydrodynamic type,
for instance KP hierarchy and its reductions, also describe
certain classes of conformal maps and curves others than analytic.
We plan to
address this question elsewhere.

{\it The $\tau$-function for curves}.
All this suggests to introduce a general notion of the $\tau$-function
for curves. For simple analytic curves, this is
a universal function which determines the curve by means of the Laurent series
\beq
\label{lg710}
x^2+y^2-t
=\sum_{k=1}^\infty{\cal R}e\Big(kt_{k}
\,(x+iy)^{k}+(x+iy)^{-k}
\frac{\p}{\p{t_k}}\mbox{log}\tau\Big)
\eeq
The dispersionless Hirota equation (\ref{Hirota}) provides a set of
differential equations
for the
$\tau$-function. Again, it has many solutions. The proper solution is
selected by the
unitarity condition (\ref{h10}) and initial data. For instance,
if the curve can
be reached by small deformations of a circle, one sets
$\p_{t_k}\mbox{log}\tau =0$  at all $t_k=0$.

{\it Moving curves and the $c=1$ string theory.}
There is another intriguing relation
(in fact equivalence) between
deformations of analytic curves and the genus-0 topological sector of 2D
gravity coupled
to
$c=1$ matter
\cite{gravity1}.  This follows from the known equivalence between the
latter and the
dispersionless Toda hierarchy restricted by the so-called
$W_{1+\infty}$-constraints
\cite{gravity1,kr2,Du2}.   The interpretation of the objects of the genus-0
string theory
in terms of analytic curves is straightforward. The positive (negative)
momentum tachyon
one-point functions $<{\cal T}_n>$ are moments of the interior of the
domain $v_n\,\,(\bar
v_n)$, the partition function of the genus-0 string is the $\tau$-function
of analytic
curves, the Schwarz function
$S(z)$ is the superpotential and, finally, the  $W_{1+\infty}$-constraints
are essentially
equivalent to the string equation (\ref{pb}) or the
relation (\ref{S4a}}). In fact, they are  nothing else than the unitarity
condition (\ref{h10}).

In its turn, the $c=1$ genus-0 string theory, as well as the relevant
solution to the
dispersionless hierarchy, is known to be equivalent to the planar limit of
the two-matrix
model (see  \cite{2matrix} and references therein).
This gives a representation of the
$\tau$-function of analytic curves
  as
$N\to
\infty$ limit of the $N\times N$ two-matrix
integral $\tau=\int DMD\bar M\exp
\mbox{tr}W$, with the potential $W=\sum_k (t_kM^k+\bar t_k\bar
M^k)+M\bar M$. We
plan to discuss  this subject elsewhere.

The interpretation of the genus-0 string theory as a simply minded
classical theory of
potential may turned out to be fruitful for both subjects. It is likely
that the
2D gravity
coupled to $c<1$ matter, being described by various reductions of the
dispersionless KP
hierarchy, also enjoys a  geometric interpretation.

{\it Laplacian growth problem.} This paper has been
stimulated by our interest in the Laplacian growth problem (see \cite{LGE}
for a review).
This problem (a source of  interesting mathematics and a great deal of
important
application) may give further insights in the matters we discuss in the
paper. It seems to
be the place for a short introduction. This is the problem of a moving
interface between
two incompressible liquids with different viscosities on the plane. Let,
say an exterior of
a  simply connected domain be occupied by a viscous fluid (oil), while less
viscous liquid
(water) occupies the interior. Water is supplied by  a source at $z=0$,
while oil  sinks
to a  source at the infinity, so the interface moves.
  Experiments and numerical
simulations  suggest that  any smooth interface, regardless of its initial
shape,
develops a finger-like pattern with a universal fractal characteristics. The
hydrodynamics of an ideal  interface (with zero surface tension)  is
described by the
Darcy law
\beq
\label{lg1a}
V_n=- \frac{\p p}{\p n}
\eeq
where $V_n$ is the velocity of the interface
and $\p p /\p n$ is the normal
derivative of the pressure on the interface. The
pressure is constant in the water domain and on the interface, while in the
oil domain it
obeys the Laplace equation $\nabla^2p=0$ with the asymptotic behavior at
the infinity
$p\to -1/2\,\mbox{log}\,
|z|$. The latter indicates a sink at the infinity.  The Darcy law
implies \cite{Richardson} that all harmonic moments
$t_k$ of the oil domain are not changed while the interface moves, but the
area of
the water domain grows linearly in time and thus can be identified with the
time.
(For connections with the inverse potential problem see \cite{EV}.) The
problem then
becomes: find the evolution of the domain as a function of the area $t$ at
fixed  moments
$t_k$.  This evolution is described by the string equation (\ref{pb}).
This equation has
a long history. It  appeared in 1945 in Ref.\,\cite{Kochina} or even
earlier  in the
mathematical theory of oil hydrodynamics. In our approach, this equation is
the basis of
the symplectic structure of the conformal maps.
Applications of integrable hierarhies to
the Laplacian growth problem are addressed in Ref.\cite{x}

At the time of completion of the manuscript
E.Fe\-ra\-pon\-tov informed us that J.Gib\-bons and S.Tsa\-rev  discussed a
relation
between Benney equations and conformal maps of slit domains
\cite{GT1}. We would like to thank J.Gib\-bons and
S.Tsa\-rev for informing us about their recent paper \cite{GT2}.

\section*{Acknowledgments}

This work has been inspired by discussions of the Laplacian growth problem
with
M.Mi\-ne\-ev-\-Wein\-stein, L.Ka\-da\-noff, L.Le\-vi\-tov and B.Shrai\-man.
We acknowledge useful dis\-cus\-sions with H.Awata, B.Dubrovin, M.Fukuma,
J.Gibbons, V.Ka\-za\-kov,  I.Kri\-che\-ver, S.P.No\-vi\-kov,
A.Or\-lov and T.Ta\-ke\-be.
Some results of this paper  were obtained in collaboration with
M.Mi\-ne\-ev-\-Wein\-stein
\cite{x}. We also thank I.Kri\-che\-ver and T.Ta\-ke\-be for teaching us
dispersionless hierarchies.

P.W. would like to thank M.Ninomiya for his hospitality
at Yukawa Institute for
Theoretical Physics in spring 1999,
where this work was started. We also thank K.Ishikawa for his
hospitality at Hokkaido
University. This work was partially supported by the Grant-in-aid for
International
Science Research (Joint Research 07044048) from the Ministry of Education,
Science, Sports and Culture, Japan. The work has been completed during
the workshop "Applications of Integrability"
at the Erwin Schr\"odinger
Institute in Vienna in September 1999.
We have been supported by grants NSF DMR
9971332 and MRSEC NSF DMR 9808595.
  A.Z. was partially supported by Russian
Foundation of
Basic Research, grant 98-01-00344.

\section*{Appendix: Ellipse growing from a circle}

Here we demonstrate how the Lax-Sato equations describe an ellipse
growing from a
circle. It is the simplest but nevertheless instructive example which in
particular allows
one to compute   the $\tau$-function at all $t_k=0$ but $t,\,t_1,\,\bar
t_1,\,t_2,\,\bar
t_2$.
Consider an ellipse with half-axes $a, b$ centered at
$z_0=x_0+iy_0$ and rotated by the angle
$\alpha$:
$$\frac{(\cos\alpha(x-x_0)-\sin\alpha
(y-y_0))^2}{a^2}+\frac{(\sin\alpha(x-x_0)+\cos\alpha
(y-y_0))^2}{b^2}=1$$
The Schwarz function of the ellipse is
$$ S(z)=e^{2i\alpha}\,\frac{a^2 +b^2}{a^2 -b^2}\,(z-z_0)+
\bar z_0-
\frac{2abe^{2i\alpha}}{a^2 -b^2}\sqrt{(z-z_0)^2-
e^{-2i\alpha }(a^2 -b^2 )}$$
The Laurent series of the Schwarz function (\ref{lg7})
$$S(z)=2t_2z+t_1+\frac{t}{z}+\frac{v_1}{z^2}+\frac{v_2}{z^3}+\ldots$$
  gives the moments of the exterior and the interior.
The only nonzero moments
   of the exterior are $t_1,\,t_2$ and their complex conjugate:
$$2t_2=e^{2i\alpha}\frac{a-b}{a+b},\;\;\;\;t_1=\bar
z_0-2t_2z_0,\;\;\;\;t=ab$$
Contrary, none of moments of the interior vanish. The first two
are
$$v_1=\frac{t(\bar t_1 +2 t_1 \bar t_2 )}{1-4t_2 \bar t_2 }\,,
\;\;\;\;
v_2=
\frac{t(\bar t_1 +2 t_1 \bar t_2 )^2}{(1-4t_2 \bar t_2 )^2}+
\frac{2t^2\bar t_2}{1-4t_2\bar t_2}$$
Adding $\p_tv_0=2\,\mbox{log}\, r$, one may check the
symmetry relations (\ref{symmetry}) and find the $\tau$-function for the
ellipse:
$$
\mbox{log}\tau=\frac{1}{2}t^2\mbox{log} t-\frac{3}{4}t^2-\frac{1}{2}t^2
\mbox{log} \,(1-4t_2\bar t_2)+t\,\frac{t_1\bar t_1+t_1^2\bar t_2+\bar t_1^2
t_2}{1-4t_2\bar t_2}
$$
Let us note that this function (for $t_1 =0$ and $t_2$ real)
was obtained \cite{FGIL}
as the free energy of a classical Coulomb gas in an external field.

The Laurent series for the conformal map
  from the exterior of the unit circle
to the exterior of the ellipse is
truncated:
$$
z(w)=rw+u_0+u_1w^{-1}\,,\;\;\;\;\;
$$
  The coefficients of the conformal map are:
$$r^2=
\frac{1}{4}(a+b)^2=
\frac{t}{1-4t_2\bar t_2}\,,
\;\;\;\;\;
u_1^2=
\frac{1}{4}e^{-4i\alpha}(a-b)^2=
\frac{4t\bar t_2^2}{1-4t_2\bar t_2}
$$
$$u_0=z_0 =\frac{\bar t_1+2t_1 \bar t_2}{1-4t_2\bar t_2}$$
The first two Hamiltonians are:
$H_1 =rw +\frac{1}{2}u_0\,$,
$H_2 = r^2 w^2 +2ru_0 w +ru_1 +\frac{1}{2}u_{0}^{2}$.
Higher flows deform the ellipse.  The  Lax-Sato equations
plus the string equation
(\ref{pb}) and their conjugate
describe how the ellipse grows from
the circle.


\begin{thebibliography}{99}

\bibitem{hydro}B.A.Dubrovin and S.P.Novikov,  Soviet Math. Dokl.
  {\bf  27} (1977) 665-669;\\
S.P.Tsarev.  Soviet Math. Dokl.
  {\bf  31} (1985) 488-491

\bibitem{kr2}
I.M.Krichever,  Funct. Anal Appl. {\bf 22} (1989) 200-213;\\
I.M.Krichever, Commun. Math. Phys. {\bf 143} (1992) 415-429;\\
I.M.Krichever, Comm. Pure. Appl. Math. {\bf 47} (1992) 437-476

\bibitem{Ta1} K.Takasaki and T.Takebe,
Rev. Math. Phys. {\bf 7} (1995) 743-808

\bibitem{gravity1}
R.Dijkgraaf, G.Moore and R.Plesser,
Nucl.Phys. {\bf B394} (1993) 356-382; \\
A Hanany, Y.Oz and R.Plesser, Nucl.Phys. {\bf B425} (1994) 150-172; \\
K.Takasaki,
Commun. Math. Phys. {\bf 170} (1995) 101-116; \\
T.Eguchi and H.Kanno, Phys.Lett. {\bf 331B} (1994) 330

\bibitem{2matrix}
J.M.Daul, V.A.Kazakov and I.K.Kostov,  Nucl.Phys.
{\bf B409} (1993) 311-338; \\
L.Bonora and C.S.Xiong, Phys. Lett. {\bf B347} (1995) 41-48

\bibitem{Brodsky}  V.Strakhov, M.Brodsky SIAM J.Appl.Math
v.46(1986),p.324-344 

\bibitem{convex}
P.S.Novikov, Soviet Math. Dokl.
  {\bf  18} (1938) 165-168;\\
M.Sakai, Proc. Amer. Math. Soc. {\bf 70} (1978) 35-38

\bibitem{GiKo1}J.Gibbons and Y.Kodama, Phys. Lett. {\bf 135 A} (1989)
167-170;\\
J.Gibbons and Y.Kodama, Proceedings of NATO ASI, 'Singular Limits of
Dispersive Waves' ed.
N.Ercolani, Plenum 1994.

\bibitem{KC}R.Carroll and Y.Kodama, 
{\bf A28}, 6373-6388 (1995)

\bibitem{AoKo}
R.Dijkgraaf and E.Witten,
Nucl. Phys. {\bf B342} (1990) 486-522;\\
A.Losev and I.Polyubin, 
 Int. J. Mod. Phys.
{\bf A10} (1995) 4161-4178;\\
S.Aoyama and Y.Kodama, 
Commun. Math. Phys. {\bf 182} (1996)
185-220

\bibitem{Du2}B.Dubrovin,
In 'Montecatini Terme 1993, Integrable systems and quantum groups' 120-348,
e-Print Archive: hep-th/9407018


\bibitem{x}M.Mineev-Weinstein, P.B.Wiegmann and A.Zabrodin,
LAUR-99-0703, submitted to Phys. Rev. Lett.


\bibitem{Davies}P.J.Davis, The Schwarz
function and its applications, The
Carus Mathematical Monographs, No. 17, The Math.
Assotiation  of America, Buffalo,
N.Y., 1974.


\bibitem{Du1}B.A.Dubrovin,
Commun. Math. Phys. {\bf 145} (1992) 195-207


\bibitem{Sakurai} M.Sakai, J. Analyse Math. {\bf 40} (1980) 144-154

\bibitem{Or}A.Orlov and E.Shulman,
Lett. Math. Phys. {\bf 12} (1986) 171-179

\bibitem{LGE}
D.~Bensimon, L.P.~Kadanoff, S.~Liang, B.I.~Shraiman, and C.~Tang,
Rev.~Mod.~Phys. {\bf 58} (1986) 977

\bibitem{Richardson} S.Richardson, J. Fluid Mech. {\bf 56} (1972) 609-618

\bibitem{EV} P.Etingof and A.Varchenko,
Why does the boundary of a round drop
becomes a curve of order four, University Lecture Series, 3. American
Mathematical
Society, Providence, RI, 1992

\bibitem{Kochina} L.A. Galin, Dokl. Akad. Nauk SSSR
{\bf 47 } (1945) 250-253;\\
P.Ya.Polubarinova-Kochina, Dokl. Akad. Nauk SSSR,
  {\bf 47 } (1945) 254-257; \\
P.P. Kufarev Dokl. Akad. Nauk SSSR
{\bf 57 } (1947) 335-348


\bibitem{GT1}J.Gibbons and S.P.Tsarev, Phys.Lett. {\bf 211A} (1996) 19-24

\bibitem{GT2}J.Gibbons and S.P.Tsarev, Phys.Lett. {\bf 258A} (1999) 263-271


\bibitem{FGIL} P.Di Franchesko, M.Gaudin, C.Itzykson and
F.Lesage,
Int. J. Mod. Phys. {\bf A9}
(1994) 4257-4351


\end{thebibliography}
\end{document}